\documentclass[arxiv, usenatbib]{iupartex-v22}
\usepackage{newtxtext,newtxmath}
\usepackage[T1]{fontenc}
\usepackage{ae,aecompl}

\usepackage{bm}		    
\usepackage{pdflscape}	

\newcommand{\qq}[1]{``#1''}

\newcommand{\rot}{\rotatebox{90}}

\newcommand{\hhsp}{\hspace*{0.2cm}}

\newcolumntype{|}{!{\vline}}
\newcolumntype{+}{>{\global\let\currentrowstyle\relax}}
\newcolumntype{^}{>{\currentrowstyle}}

\newcommand{\an}{{Astr. Nachr.}}
\newcommand{\parep}{PAR}

\newcommand{\basi}{Bull. Astron. Soc. India}
\chardef\us=`\_

\newcommand{\tb}[1]{\textbf{#1}}

\title[29 March 2006 Solar Corona]{Physical Properties of 29 March 2006 Solar Corona}

\author[H. \c{C}akmak]{%
H. \c{C}akmak$^{1\cc}$\orcid{0000-0002-1959-6049}
\affsep \\
$^1$\.Istanbul University, Faculty of Science, Department of Computer Sciences, 34116, Beyaz\i t, \.Istanbul, T\"urkiye
}

\corres{H. \c{C}akmak}{hcakmak@istanbul.edu.tr}

\pubyear{2025}

\doiheader{XXXXXXX/PAR.20XX.00000}
\date{
	\pSubmit{18.07.2025} 
	\pRevReq{10.09.2025}
	\pLastRevRec{28.09.2025}
	\pAccept{08.10.2025}
	\pPubOnl{XX.XX.2025}
}
\volume{0}
\volnumber{1}

\begin{document}
\label{firstpage}
\pagerange{\pageref*{firstpage}--\pageref*{lastpage}}
\maketitle

\begin{abstract}
On 29 March 2006, a total solar eclipse was observed in the Manavgat district of Antalya, Turkey. During the event, the solar corona was observed using an 8-inch mirrored telescope. White-light polarization observations were carried out at three distinct angles using a polarizing filter placed in front of the camera system. To calibrate the intensity of the roll film, photographs of the eclipse and the solar disk were taken with a traditional 35mm manual camera. Using the solar disk images obtained during the eclipse, an intensity calibration curve for the roll film was created. This curve was then used to calculate various physical properties of the solar corona, including intensity, degree of polarization, electron density, and mean temperature. The results of these calculations were compared with the corona models developed by \cite{VDH1950} and \cite{SK1970}, as well as with findings from other researchers. Except for the degree of polarization, the measured physical parameters closely match the values given in the literature.
\end{abstract}



\section{Introduction}\label{sec:1-intro}
The primary importance of total solar eclipses lies in the opportunity they provide to observe the solar corona during the brief period of totality. The solar corona is defined as the Sun's enigmatic outer atmosphere, which is characterized by its extreme faintness and is approximately one million times dimmer than the Sun's photosphere. This characteristic makes it challenging to observe directly, except during a solar eclipse. Despite its low particle density, the corona has an incredibly high temperature, reaching several million degrees. This paradox, in addition to other unresolved questions, makes every total solar eclipse a valuable scientific event that draws significant attention. The corona's appearance, or morphology, suggests that its primary source of energy is the Sun's magnetic field. Changes in the strength and structure of the magnetic field, particularly between different phases of the solar cycle, affect the corona's visible shape. During the solar cycle's minimum phase, coronal structures tend to concentrate in the area near the Sun's equator. However, during the maximum phase, these structures exhibit a more even distribution across the entire solar disk. This finding suggests a close correlation between the shape and intensity of the solar corona and the activity of the sunspot cycle.

The appearance of the solar corona during the minimum period is characterized by a nearly uniform gradient in intensity, extending from the nearest regions of the solar disk to the distant regions. This smooth gradient has made the development of a relatively simple model for minimum type corona possible, depending on the observed coronal intensity and the distance to the solar disk edge. The first model of the coronal intensity and polarization was developed by \cite{SA1879}, who provided the main solutions to all fundamental mathematical problems in this subject. Subsequently, \cite{MM1930} advanced the Schuster theory by including the limb-darkening effect of the solar disk, thereby enhancing its complexity and precision. Subsequently, the seminal works of \cite{BS1937, BS1939} advanced this field by introducing the first general electron density function in a polynomial function with the form of $r^{{\rm -n}}$, developed using white-light corona data obtained photometrically.

In the present day, the observation of total solar eclipses has a significant role in the verification of solar corona models. This is because they provide valuable insights into the general condition of the corona during any sunspot cycle. Following the detailed characterization of the general conditions in the solar corona for the minimum type of solar cycle, the variations from one solar cycle to another will be more effectively identified. This will provide an opportunity to specify the variations that have occurred during any sunspot cycle. Each solar corona exhibited during a total solar eclipse is characterized by distinct properties, which vary from one another to a certain extent. The total solar eclipse of 29 March 2006 occurred close to the time of the minimum phase of the 23rd solar cycle. This event is of particular importance as it provided further observational results for testing the developed corona models.

This study is an extensive review of the observational results obtained for the 2006 eclipse. It is evident from the results obtained that the solar corona observed during this eclipse is consistent with the characteristics of the minimum type of corona. All physical parameters, with the exception of polarization, demonstrate a high degree of agreement with the models and numerous observational values as reported in the current literature. The instruments utilized for observation and data reduction are delineated in Section~\ref{sec:2-obsdat}, while the stages of intensity calibration performed are explained in Section~\ref{sec:3-intcal}. The subsequent Section~\ref{sec:4} presents the calculated values of total coronal ($K+F$) intensity and polarization degree, along with the $K$ coronal intensity and its polarization degree $P_{\rm K}$ values. Furthermore, the results of electron density and average coronal temperature are given in Sections~\ref{sec:5} and \ref{sec:6}, respectively. Finally, a general discussion of the results of the eclipse is provided in Section~\ref{sec:7}.
\begin{figure*}[h!]   
	\centerline{\includegraphics[width=0.78\linewidth]{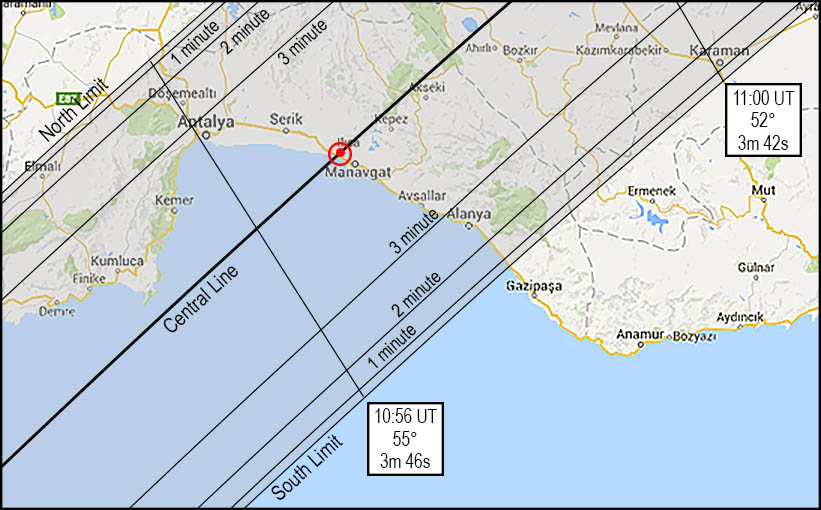}} \vspace{0.3ex}
	\centerline{\includegraphics[width=0.7825\linewidth]{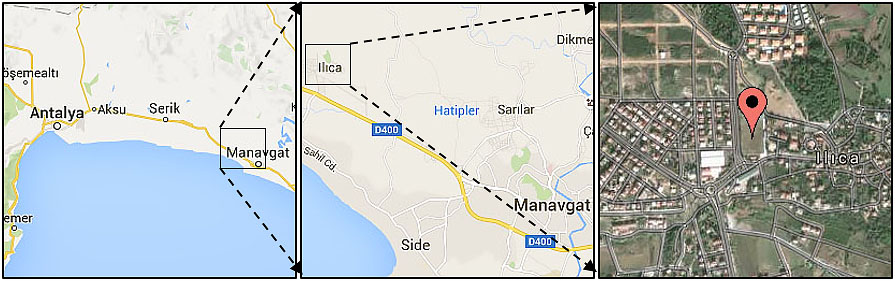}}
	\caption{The upper panel illustrates the position of the total eclipse cone of 29 March 2000 over the southern part of Turkey. The location of the observation site is indicated by a small red point and a circle. This site is situated in close proximity to the central line of the eclipse. Furthermore, the location of the observation site at Ilıca City Stadium is denoted by a small red location icon, as illustrated in the bottom-right image.} \label{fig:1}
\end{figure*}

\section{Observation and Data Reduction}\label{sec:2-obsdat}
The observation of the solar corona during the total solar eclipse on 29 March 2006 was conducted by the staff of the Astronomy and Space Sciences Department of Istanbul University in Ilıca, a town in the Manavgat district of Antalya, Turkey. The site of observation has coordinates of $31^\circ 22^\prime$ E and $36^\circ 49^\prime$ N, and an altitude of 42 meters. As illustrated in the upper panel of Figure~\ref{fig:1}, the observation location is proximate to the center line of the totality, about 650 m from the center line. Furthermore, certain circumstances associated with this eclipse are given in Table~\ref{tab:1}.

On the day of observation, weather conditions were favourable for the eclipse, with the sky being clear and free of clouds or fog. Consequently, a series of observations was conducted simultaneously at the observation site. Each observation had a specific purpose, and different types of equipment were used. The white-light polarization observation selected for this study was one of the observations made. The polarization observation was conducted using an 8-inch Meade brand mirrored telescope, which possesses a 203mm aperture and 1280mm focal length, with a classic 35mm manual camera attached to its focal point. A polarizing filter with a transmittance of 30\% was utilized, which was then incorporated into an apparatus capable of rotation at three distinct angles $0^\circ,  60^\circ$, and $120^\circ$, attached to the front of the camera.
\begin{table*}[h!]
	\setlength{\tabcolsep}{4.7pt}
	\renewcommand{\arraystretch}{0.9}\centering\small
	\caption{Local circumstances of the March 29, 2006 eclipse at Ilıca site }\label{tab:1}
	\begin{tabular}{p{0.5cm}ll}
		\hline\hline
		\multicolumn{2}{l}{Eclipse parameters} &  \\ [-0.5ex]
        \hline
		& Beginning of totality & $10^h \; 55^m \; 02^s$  UT \\
		& End of totality & $10^h \; 58^m \; 47^s$  UT \\
		& Duration of totality & $3^m \; 45^s$ \\
		& Width of totality & 170 km \\
		& Altitude of the Sun during totality & from 56$^\circ$ to 44$^\circ$ \\
		& Ratio of the radii of apperent Sun and Moon   & 1.049 \\
		& Position angle of second contact & 47$^\circ$ $\,$ from celestial N to E\\
		& Position angle of third contact & 228$^\circ$ from celestial N to E\\
		&   &  \\
		\multicolumn{3}{l}{Parameters of the Sun's rotation axis and its equator} \\ [-0.5ex]
        \hline
		& Position angle, \textit{P} (from celestial N to W)  & - 26$^\circ$.02  \\
		& Equator angle, $B_0$ (latitude of apparent Sun's center)&  - $\,$ 6$^\circ$.68 \\
		\hline\hline
	\end{tabular}
\end{table*}
\begin{figure*}[h!]   
	\centerline{\includegraphics[width=0.85\textwidth, clip=]{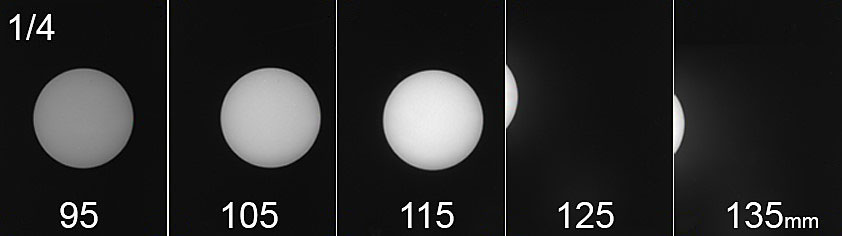}}
	\caption{Here is an example of the solar disk images taken for intensity calibration at the 1/4 second exposure with 5 different diaphragm openings. In the last two images, most of the solar disk has been moved outside the frame to prevent excessive saturation of the sky background}\vspace*{-3pt}
     \label{fig:2}
\end{figure*}
\def\mba{\hspace{44pt}}
\def\mbb{\hspace{30pt}}
\def\msa{\hspace{25pt}}
\def\mbc{\hspace{44pt}}
\def\msb{\hspace{22pt}}
\def\msc{\hspace{22pt}}
\def\msd{\hspace{12pt}}
\def\mbx{\hspace{19pt}}
\def\mby{\hspace{21pt}}
\def\mse{\hspace{32pt}}
\def\fgn{0.81\linewidth}
\begin{figure*}[h!]    
	\small
	\centerline{\includegraphics[width=\fgn]{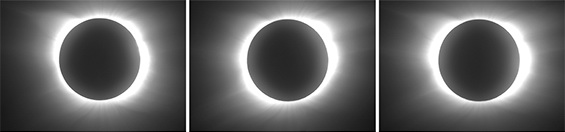}} \mba 1/2 
	\mbb $11^h \; 55^m \; 10^s$ \msa $0^\circ$
	\mbc $11^h \; 55^m \; 17^s$ \msb $60^\circ$
	\mbc $11^h \; 55^m \; 23^s$ \msd $120^\circ$ \\[-6pt]
	
	\centerline{\includegraphics[width=\fgn]{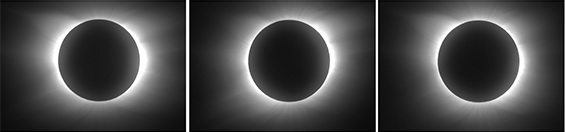}} \mba 1/4 
	\mbb $11^h \; 55^m \; 41^s$ \msa $0^\circ$
	\mbc $11^h \; 55^m \; 35^s$ \msb $60^\circ$
	\mbc $11^h \; 55^m \; 28^s$ \msd $120^\circ$ \\[-6pt]
	
	\centerline{\includegraphics[width=\fgn]{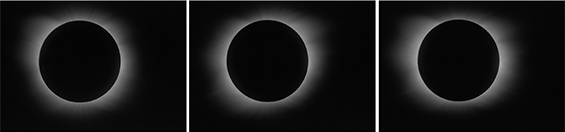}} \mba 1/30
	\mbx $11^h \; 56^m \; 22^s$ \mse $0^\circ$
	\mbc $11^h \; 56^m \; 29^s$ \msb $60^\circ$
	\mbc $11^h \; 56^m \; 35^s$ \msd $120^\circ$ \\[-6pt]
	
	\centerline{\includegraphics[width=\fgn]{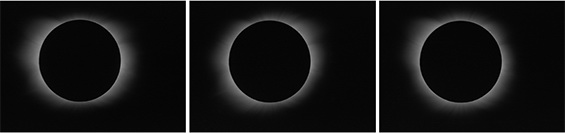}} \mba 1/60 
	\mbx $11^h \; 56^m \; 53^s$ \mse $0^\circ$
	\mbc $11^h \; 56^m \; 47^s$ \msb $60^\circ$
	\mbc $11^h \; 56^m \; 40^s$ \msd $120^\circ$ \\[-6pt]
	
	\centerline{\includegraphics[width=\fgn]{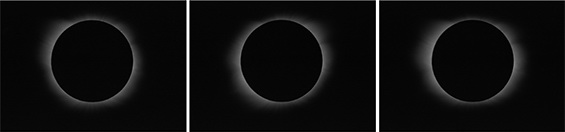}} \mba 1/125 
	\mby $11^h \; 56^m \; 58^s$ \msa $0^\circ$
	\mbc $11^h \; 57^m \; 05^s$ \msb $60^\circ$
	\mbc $11^h \; 57^m \; 11^s$ \msd $120^\circ$ \\[-10pt]
	\normalfont
	\caption{All the images of the eclipse that were utilized in the calculation process. The sequence of numbers from left to right on the bottom of the image corresponds to the exposure time, the time of observation, and the polarization angle, respectively.}\vspace*{-5pt}
    \label{fig:3}
\end{figure*}
Before observation, 25 photographs of the solar disk were taken using a solar filter with a transmittance of $16\times 10^{-6}$ in order to calibrate the intensity response of the roll film used during the experiment. To avoid image saturation and ensure a broad dynamic range in the captured images, a combination of five diaphragm apertures with 95, 105, 115, 125, and 135mm in diameter was employed along with five different exposure times: 1/2, 1/4, 1/30, 1/60, and 1/125 seconds. An example set of these solar disk images, along with the corresponding diaphragm diameters, is presented in Figure~\ref{fig:2}. During totality, 36 photographs were taken over a period of $3^{\rm m} 45^{\rm s}$ using a polarizing filter set at three different polarization angles. The same five exposure times used in the calibration shots were applied during this phase as well. From this collection, 15 images were selected for the analysis of the solar corona’s physical properties. These selected photographs, along with their associated exposure times and polarization angles, are shown in Figure~\ref{fig:3}.

As seen in Figure~\ref{fig:3}, the use of a wide-aperture telescope with 35mm film has resulted in an asymmetric space around the solar disk on the images, with the space on the polar regions being smaller than that on the equatorial regions. After the eclipse, the film was developed in the laboratory of the Department of Astronomy and Space Sciences, Faculty of Science, Istanbul University. All the eclipse and calibration images were then scanned into a computer using a Microtek Artix Scan 4000t Film Scanner, which features a 16-bit dynamic range, a pixel intensity range of 0 to 65535, and a 2000 DPI resolution.

\section{Intensity Calibration}\label{sec:3-intcal}
Intensity calibration of the images utilized in the observational stage constitutes a crucial phase within the post-eclipse procedures. Despite the production of an intensity-exposure curve by the film manufacturer, this is frequently insufficient for the observation in question. It is acknowledged that the intensity observed on the images varies in every eclipse observation. This is because the weather conditions and types of equipment used can have a significant impact on the results obtained. Consequently, the intensity-exposure curve of the images should be reformed for the current active observation. In the case of the 2006 eclipse, solar disk images captured with a solar filter were utilized to establish the \qq{intensity calibration function} (ICF) of the images, employing a novel approach that had been developed. A comprehensive and detailed explanation of this topic can be found in \cite{HC2017}. This article provides a concise overview of the processes to be performed, delineating two primary definitions. The first of these is the normalized intensity, I$_N$, which is employed to express any intensity in an image in terms of background intensity. This is given by
\begin{equation}\label{equ:1}
	I_{{\rm N}} = \frac{I_0}{I_{\rm min}}
\end{equation}
where I$_0$ is the intensity measured on the images and I$_{\rm min}$ is the lowest background intensity among all exposures taken.

The second one, the relative intensity, $I_{\rm R}$, is employed to express the observed intensity in the unit of average solar disk brightness. The efficacy of this parameter is dependent upon the equipment utilized during observation, including the telescope aperture, the diaphragm, the transmittance of the polarize and solar filter, and the exposure time. In consideration of the assertions presented in the publications of \cite{MWR1994} and \cite{AMK2007}, the relative intensity $I_{\rm R}$ can be expressed with the aforementioned parameters as,
\begin{equation}\label{equ:2}
	I_R = \frac{I_{\rm obs}}{I_{\rm \odot}} = f_{\rm int} * f_{\rm pol} * t * \bigg( \frac{A_{\rm D}}{A_{\rm O}} \bigg)
\end{equation}
where I$_{obs}$ is the observed intensity. The quantity $I_{\odot}$ is average solar disk brightness, besides $f_{\rm int}$ and $f_{\rm pol}$ represent the transmittance of the solar filter and polarizer, respectively. The exposure time is denoted by $t$ and the radii of the diaphragm opening and telescope aperture are indicated by $A_{\rm D}$ and $A_{\rm O}$, respectively. The intensity calibration function (ICF) is then derived through the application of an exponential curve, shown as $y=A\times e^{Bx}$, to the graphical representation formed between the $I_{\rm N}$ and I$_{\rm R}$ values.

\begin{figure}[t]
	\centerline{\includegraphics[width=0.52\linewidth]{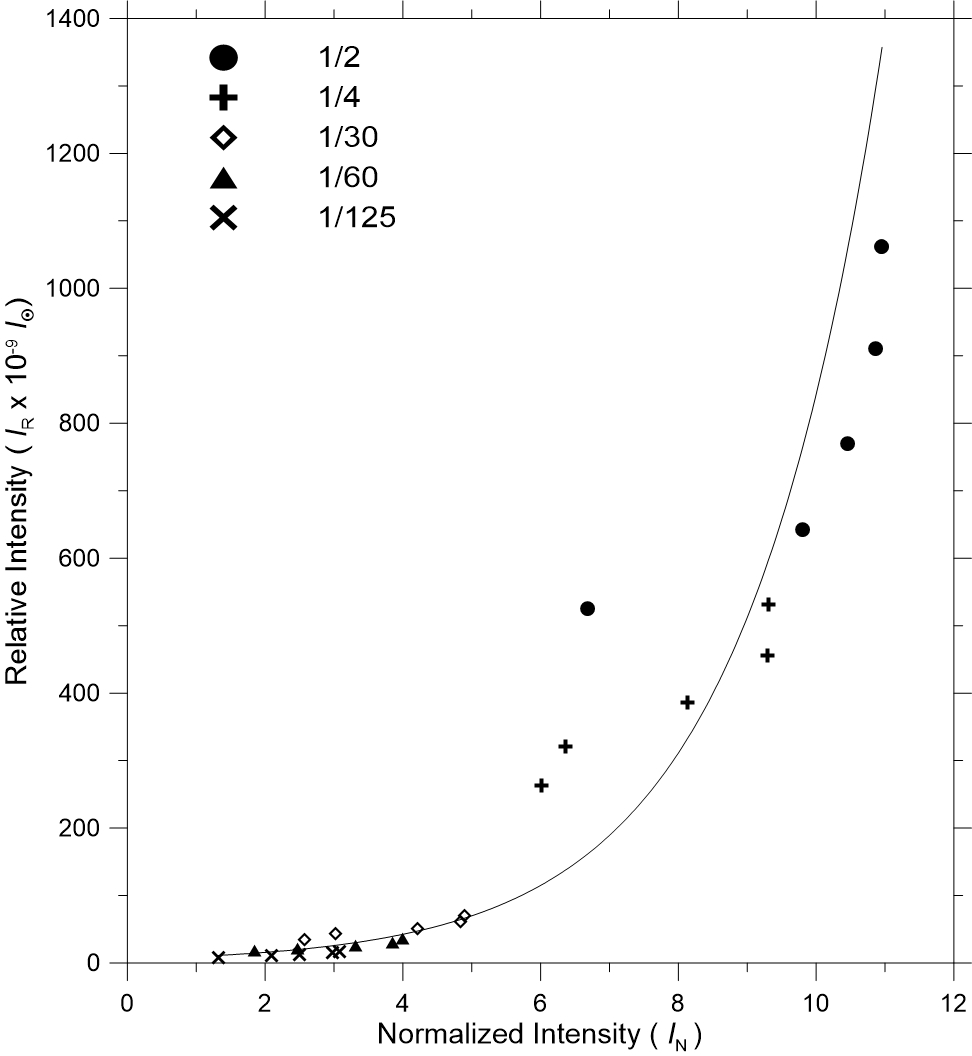}}
	\caption{The intensity calibration curve acquired for the 2006 eclipse. Each exposure time is shown using a different symbol. The solid line represents the curve fitted.}
    \label{fig:4}
\end{figure}

As elucidated above, the mean intensity of each solar disk in the calibration images is measured as the mean intensity of a specific area over the disk that contains almost no saturated pixels. Due to the shifted disk images, in all images, the same region near the edge of the disk was selected for solar chromospheric disk brightness. Subsequently, the mean intensity of each calibration image is divided by the lowest background intensity observed in all exposures. The lowest value recorded was 5874. In addition, the transmittance of the solar filter and polarizer utilized in this eclipse observation is, respectively, $16\times 10^{-6}$ and 0.3. The calculated values of normalized and relative intensity are enumerated in Table~\ref{tab:2}, alongside the exposure times and diaphragms utilized. Figure~\ref{fig:4} presents a graph between $I_{\rm N}$ and $I_{\rm R}$ values, with the curve fitted. The fitted curve function is obtained as
\begin{equation}\label{equ:3}
	I_{\rm R} = 5.7969 \times 10^{-9} \hspace*{0.1cm} e^{0.4979 \hspace*{0.05cm} I_{\rm N}}
\end{equation}
with a correlation coefficient $R^2=0.97$. This function provides the relative intensity for a given normalized intensity on the images utilized in this eclipse.
\begin{table*}[t]
	\setlength{\tabcolsep}{10pt}
	\renewcommand{\arraystretch}{1}
	\setlength{\extrarowheight}{2pt}
	\centering\small
	\caption{Normalised and relative density values calculated using calibration images from the 2006 eclipse.}
	\begin{tabular}{p{0.6cm}lrrrrr}
    \hline
    \hline
		& & \multicolumn{5}{c}{Diameter of diaphragm opening in mm} \\
		\cline{3-7}
		& \multicolumn{1}{c}{Exposure (s)}  & \multicolumn{1}{c}{95} & \multicolumn{1}{c}{105} & \multicolumn{1}{c}{115} & \multicolumn{1}{c}{125} & \multicolumn{1}{c}{135} \\
		\cline{2-7}
		\multicolumn{1}{c}{\multirow{5}[2]{*}{\rot{\parbox[t][0.01\linewidth][t]{0.1\textwidth}
					{\centering Normalized\\Intensity}
		}}} 
		& \hhsp 1/2 & 6.674 & 9.799 & 10.459 & 10.871 & 10.957 \\
		\multicolumn{1}{c}{} & \hhsp 1/4 & 6.016 & 6.364 & 8.132 & 9.292 & 9.314 \\
		\multicolumn{1}{c}{} & \hhsp 1/30 & 2.576 & 3.019 & 4.214 & 4.834 & 4.899 \\
		\multicolumn{1}{c}{} & \hhsp 1/60 & 1.853 & 2.472 & 3.315 & 3.850 & 3.996 \\
		\multicolumn{1}{c}{} & \hhsp 1/125 & 1.318 & 2.097 & 2.494 & 2.983 & 3.078 \\
		\cline{2-7}
		&       &       &       &       &       &  \\[-10pt]
		\cline{2-7}
		\multicolumn{1}{c}{
			\multirow{5}[3]{*}{\rot{\parbox[b][0.01\linewidth][t]{0.1\linewidth}
					{\centering Relative Int.\\($\times 10^{-9} \bar{I}_{\odot}$)}
		}}} 
		& \hhsp 1/2 & 525.613 & 642.093 & 770.220 & 909.995 & 1061.419 \\
		\multicolumn{1}{c}{} & \hhsp 1/4 & 262.807 & 321.046 & 385.110 & 454.998 & 530.709 \\
		\multicolumn{1}{c}{} & \hhsp 1/30 & 35.041 & 42.806 & 51.3480& 60.666 & 70.761 \\
		\multicolumn{1}{c}{} & \hhsp 1/60 & 17.520 & 21.403 & 25.674 & 30.333 & 35.381 \\
		\multicolumn{1}{c}{} & \hhsp 1/125 & 8.410 & 10.273 & 12.323 & 14.560 & 16.983 \\
        \hline
        \hline
	\end{tabular}
    \label{tab:2}
\end{table*}

After this step, the normalization process of the eclipse images is conducted by dividing the intensities of the solar corona by the lowest background intensity among all images of the eclipse. Subsequently, the relative intensity values for these normalized intensities of the corona in images are calculated by using the ICF given in Equation(\ref{equ:3}). Furthermore, it is important to note that another utility of this ICF is that the instrumental and sky contributions (I$_{\rm A+S}$) to the total intensity can be found by taking the normalized intensity to be equal to 1 in Equation(\ref{equ:3}). This value ($I_{\rm N}=1$) is representative of the intensity of the sky itself. For this eclipse, this value is calculated as 0.95$\times 10^{-8} I_\odot$, which is relatively close to the value of 0.5$\times 10^{-8} I_\odot$ obtained by \cite{KK2005} during the eclipse of August 11, 1999. However, this value is slightly larger than the value of 0.2$\times 10^{-8} I_\odot$ obtained by \cite{NEP1961} during the eclipse on 2 October 1959.

\section{Brightness and Polarization}\label{sec:4}
Before the calculations of brightness and polarization, it is necessary to create a composite image for each polarisation angle taken at five different exposure times that are 1/2, 1/4, 1/30, 1/60, and 1/125 seconds, respectively. The intensity in the composite image must be the average of the intensities taken in different exposures. In this case, since there are five different exposures, the sum of the intensity values in each exposure must be divided by five. In addition, due to the linear correlation between intensity and exposure time, the total exposure times should also be taken into account in the intensity calculation \citep{HC2017}. In this observation, the composite intensity in each of the polarization angles ($0^\circ, 60^\circ$, $120^\circ$) is calculated using an empirical formula developed as
\begin{equation}\label{equ:4}
	I_{\rm C} = \frac{1}{\Sigma t_{\rm exp}} \frac{I_{1/2} + I_{1/4} + I_{1/30} + I_{1/60} + I_{1/125}}{5}
\end{equation} 
where $\Sigma t_{\rm exp}$ denotes the sum of selected exposure times in seconds and written as
\begin{equation}
	1/2 + 1/4 + 1/30 + 1/60 + 1/125 = 0.808 \; \text{second} \nonumber
\end{equation}
where $I_{1/2}$ denotes the measured intensity of the image exposed for 1/2 second, $I_{1/4}$ denotes exposure of 1/4 second, and so forth. By utilizing these specific intensities of each polarization angle, the Stokes parameters for this eclipse observation are obtained by
\begin{equation}\label{equ:5}
	\begin{split}
		I &= \frac{2}{3} \big( I_{\rm 0} + I_{\rm 60} + I_{\rm 120} \big) \\
		Q &= \frac{2}{3} \big( 2\, I_{\rm 0} - I_{\rm 60} - I_{\rm 120} \big) \\
		U &= \frac{2}{\sqrt{3}} \big( I_{\rm 60} - I_{\rm 120} \big), \quad V = 0
	\end{split}
\end{equation}
where $I$ is the total coronal brightness ($K+F$), $Q$ is the amount of linear polarization in the vertical or horizontal plane, $U$ is the value on $+45^\circ$ or $-45^\circ$ plane, and $V$ is the amount of circular polarization which is assumed to be equal to zero in the eclipse observations \citep{BDE1966, GD2003}. Subsequently, as expressed in \cite{GD2003}, the degree of polarization, denoted $P$, and the angle of polarization, denoted by theta, are calculated for the total intensity by
\begin{equation}\label{equ:6}
	\begin{split}
		P &= \sqrt{Q^2 + U^2} \\
		\theta &= \frac{1}{2} \; {\rm arctan} \bigg( \frac{U}{Q} \bigg)
	\end{split}
\end{equation}

\begin{figure}[t]   
	\centerline{\includegraphics[width=0.49\linewidth]{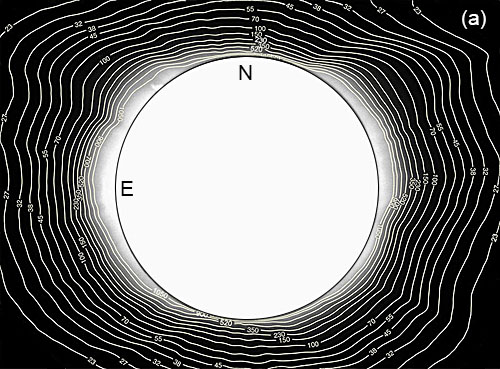} \hspace{1pt}\includegraphics[width=0.49\linewidth]{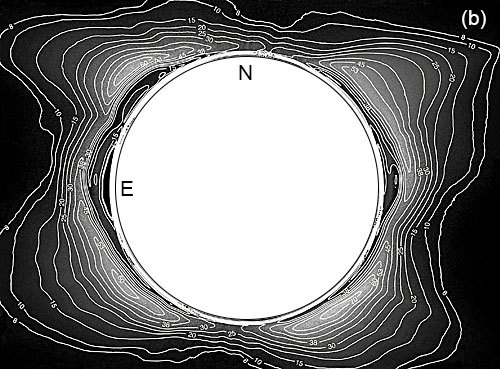}}
	\caption{\tb{(a)} Isophotes of total coronal brightness (numbers are in the unit of 10$^{-9} I_{\odot}$) and \tb{(b)} isolines of polarization degree (numbers are in percent) of the solar eclipse March 29, 2006.}
     \label{fig:5}
\end{figure}

After subtracting the instrumental and sky contribution values $I_{\rm A+S}$ from the observed total coronal density, the total coronal density $K+F$ and polarization degree $P_{\rm K+F}$ values were calculated for the 2006 eclipse, and the resulting images are shown in Figures~\ref{fig:5}a and ~\ref{fig:5}b, respectively. As illustrated in both figures, the selection of appropriate isolines for intensity and polarization degree was carefully performed. It is important to note that particular attention was paid to ensure that the selected isolines demonstrate the general gradient over a wide range and are equally spaced. The total corona intensities are calculated in all radial directions of polar angles between 0 and 360 degrees, with 5-degree steps. As demonstrated in Table~\ref{tab:3}, the values of the four special polar angles ($0^\circ$, $30^\circ$, $60^\circ$, and $90^\circ$) and the average total intensity values of the equatorial and polar regions are listed. In addition, the average polarization degree values of the equatorial and polar regions are also presented.
\begin{table}[t]
	\setlength{\tabcolsep}{8pt}
	\renewcommand{\arraystretch}{0.7}
	\setlength{\extrarowheight}{5pt}
	\centering \small	
	\caption{Observational total coronal brightness $K+F$ and polarization degree ($P_{\rm K+F}$) values of 29 March 2006 eclipse.}
	\begin{tabular}{crrrrrrrrr}
    \hline
    \hline
		& \multicolumn{6}{c}{$K+F$ corona ($\times 10^{-9} I_{\odot}$)} &  & \multicolumn{2}{c}{$P_{\rm K+F}$ (\%)} \\
		\cline{2-7}\cline{9-10}
		$r(r_\odot)$ & \multicolumn{1}{c}{0} & \multicolumn{1}{c}{30} & \multicolumn{1}{c}{60} 
		& \multicolumn{1}{c}{90} & Equ. & Pol. &  & Equ. & Pol. \\
		\cline{1-7}\cline{9-10}
		1.10 &  510.9 & 891.0 &       &       & 1310.9 & 648.2 &  & 19.03 & 32.39 \\
		1.15 &  280.8 & 570.1 & 984.2 &       & 1046.3 & 364.2 &  & 26.73 & 36.30 \\
		1.20 &  166.4 & 308.7 & 732.2 & 983.5 &  781.8 & 200.5 &  & 36.44 & 33.17 \\
		1.25 &  110.6 & 174.9 & 468.1 & 701.2 &  513.2 & 128.1 &  & 44.69 & 25.95 \\
		1.30 &   81.2 & 118.2 & 265.4 & 432.4 &  292.8 & 91.0  &  & 46.12 & 19.57 \\
		1.35 &   64.4 &  86.6 & 172.6 & 251.6 &  180.2 & 70.1  &  & 39.79 & 15.00 \\
		1.40 &   53.8 &  66.9 & 125.8 & 159.6 &  125.4 & 57.2  &  & 32.40 & 13.87 \\
		1.45 &   45.9 &  53.9 &  97.7 & 111.8 &  94.0  & 48.1  &  & 26.81 & 12.28 \\
		1.50 &        &  45.1 &  78.8 &  84.6 &  74.3  & 42.3  &  & 22.47 & 11.56 \\
		1.55 &        &  39.3 &  65.8 &  67.9 &  61.2  & 37.7  &  & 19.40 & 10.67 \\
		1.60 &        &  34.7 &  56.6 &  56.2 &  52.0  &       &  & 17.11 & \\
		1.65 &        &  30.4 &  49.6 &  47.8 &  45.1  &       &  & 15.84 & \\
		1.70 &        &       &  44.0 &  41.7 &  39.9  &       &  & 14.61 & \\
        \hline
        \hline
	\end{tabular}
    \label{tab:3}
\end{table}
\begin{figure}[h!]   
	\begin{minipage}[b]{.48\linewidth} 
	\centerline{\includegraphics[width=0.9\linewidth]{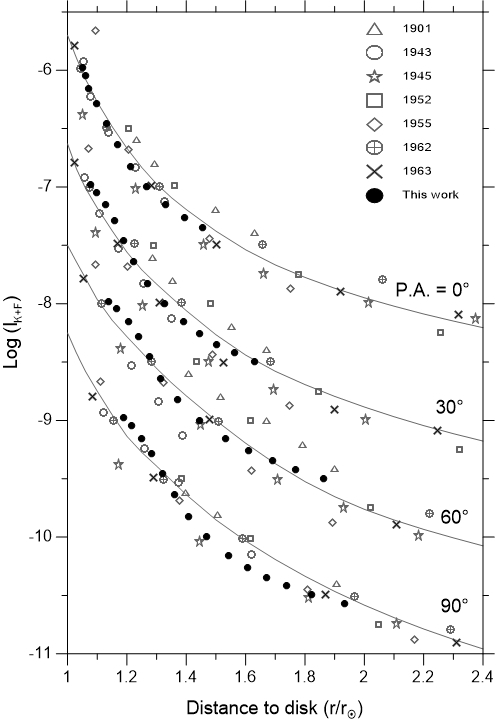}}
	\caption{The average total coronal brightness of the solar eclipse on 29 March 2006 is shown at four specific polar angles, in comparison with the model values of \protect\cite{SK1970} (straight line) and the observational values of seven eclipses (shown as symbols).} 
    \label{fig:6}
\end{minipage}\hspace*{11pt}
\begin{minipage}[b]{.48\linewidth} 
	\centerline{\includegraphics[width=0.99\linewidth]{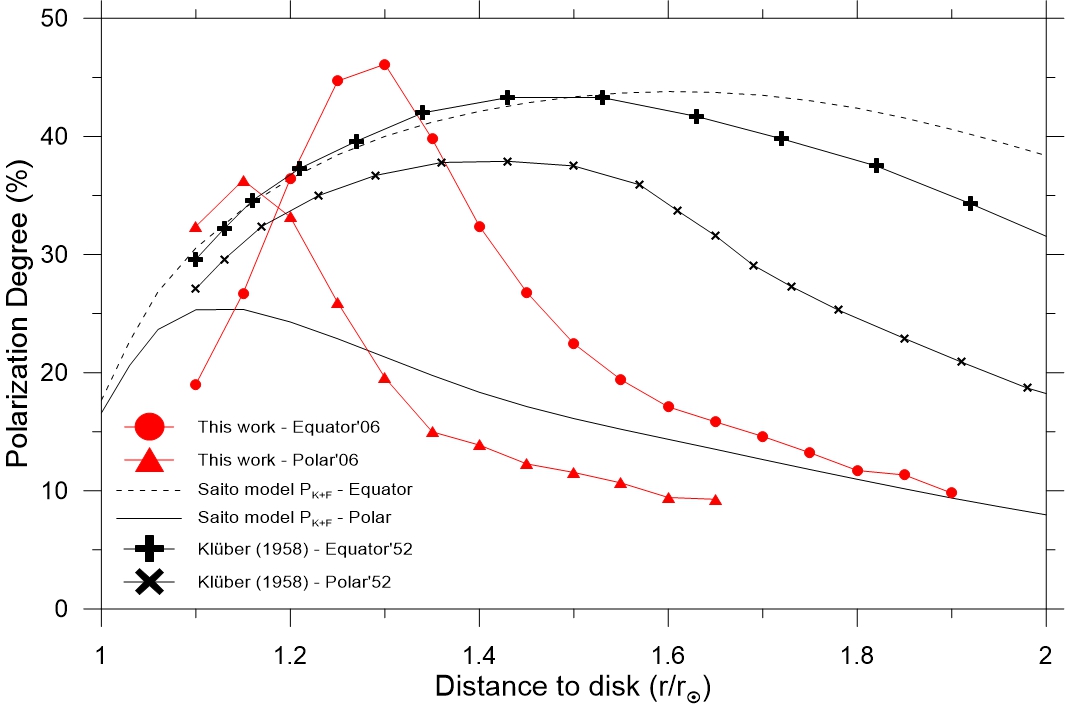}}
	\caption{The polarization degree $P_{{\rm K+F}}$ of the solar eclipse on 29 March 2006 is compared with the model values of the polarization degree of \protect\cite{SK1970} (the line with no symbol) and the observational results of \protect\cite{KH1958}.}
    \label{fig:7}\vspace*{11pt}
\end{minipage}
\end{figure}

Figure~\ref{fig:6} shows the intensity values separately for the polar angles of four specific directions, compared with the model values of \cite{SK1970} and the observational values of seven eclipses \citep{YRK1911, SK1948, RJM1951, SK1956, KH1958, SH1964, WM1964}. In the figure, the values for the $30^\circ$ polar angle are shifted downwards by 1 unit, the values for the $60^\circ$ polar angle by two units, and the values for the $90^\circ$ polar angle by three units, so that all values can be seen clearly. Therefore, the numbers on the y-axis below -6 do not represent the true values for the polar angles of $30^\circ$, $60^\circ$, and $90^\circ$. As can be seen in Figure~\ref{fig:6}, the total intensity values obtained for the 2006 eclipse show quite good agreement with both the model and other observational values. The fluctuations in total coronal brightness depending on distance from the disk are particularly evident at $60^\circ$ and $90^\circ$ polar angles and are thought to be mainly caused by an asymmetric distribution of material in the corona due to changes in magnetic activity depending on the phase of the solar cycle. Furthermore, this asymmetrical appearance is clearly visible in the polarization degree shown in Figure~\ref{fig:5}a.
During the calculation of the average total brightness, the polar angles of $260^\circ$, $270^\circ$, $280^\circ$, and $310^\circ$, at which active coronal structures or regions exist, were not taken into account to obtain a general intensity distribution and make a more appropriate comparison with the model values. This ensures spherical symmetry and an isotropic distribution of the observed intensity values.
\begin{figure}[t]
	\begin{minipage}[hb]{.48\linewidth}    
		\centerline{\includegraphics[width=.99\textwidth]{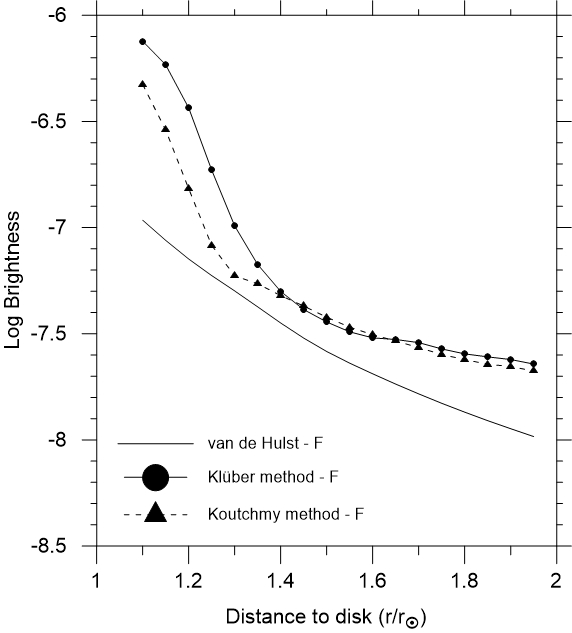}}
		\captionsetup[figure]{labelsep=space}
		\captionof{figure}{The observational $F$ corona values obtained by two different methods are compared with the model values of \protect\cite{VDH1950}.}
	    \label{fig:8}
	\end{minipage}\hspace*{11pt}
	\begin{minipage}[ht]{.48\linewidth} 
		\vspace*{-90pt}\setlength{\tabcolsep}{10pt}
		\renewcommand{\arraystretch}{1.1}\centering
		\captionsetup[table]{labelsep=space}
		\captionof{table}{The $K$ coronal brightness values in $10^{-9} I_{\odot}$ unit for the equatorial and polar regions of the 2006 eclipse.}\small \label{tab:4}\vspace*{5pt}
		\begin{tabular}{crr}
			\cline{1-3}
			r($r_\odot$)  & Equ.  & Polar \\
			\cline{1-3}
			1.10 &       & 547.8 \\
			1.15 & 965.7 & 284.9 \\
			1.20 & 716.2 & 136.8 \\
			1.25 & 459.9 &  75.6 \\
			1.30 & 249.1 &  47.1 \\
			1.35 & 143.9 &  32.9 \\ 
			1.40 &  93.9 &  25.2 \\
			1.45 &  66.4 &  21.0 \\
			1.50 &  49.9 &  10.3 \\
			1.55 &  39.5 &       \\
			1.60 &  32.4 &       \\ 
			1.70 &  23.9 &       \\
			1.80 &  19.2 &       \\
			\cline{1-3}
		\end{tabular}
	\end{minipage}
\end{figure}
The polarization degree $P_{\rm K+F}$ of the total coronal brightness of the 2006 eclipse is shown in Figure~\ref{fig:7} in comparison with the model values of \cite{SK1970} and the results of the 25 February 1952 eclipse of \cite{KH1958}. The 1952 eclipse occurred at a relatively similar stage of the solar cycle to that of the 2006 eclipse. The general trend of the polarization degree of this eclipse seems to be quite different from that of the model and other observations. But similar trends in the degree of polarization can also be seen in the results of observations made by \cite{JJJ1934} at the eclipse on 14 February 1934, \cite{FB1935} at the eclipse on 21 August 1914, \cite{DG1936} at the eclipse on 31 August 1932, \cite{AM1970} at the eclipse on 12 November 1966, \cite{KS1978} at the eclipse on 30 June 1973 and \cite{RA1986} at the eclipse on 16 February 1980. 

A comparison of the obtained maximum values of polarization degree with \cite{SK1970} model values reveals the following conclusions. The maximum value of 46\% in the equatorial region is reached at a distance of approximately $1.3 R_\odot$, while the maximum value of the model in the equatorial region is 44\% at a distance of $1.6 R_\odot$. As can be seen from these values, both maxima are quite close to each other, but there is a $0.3 R_\odot$ difference between the distances at which the maxima occur. The same situation is observed in the polar region. The maximum observational value is 36\% at a distance of $1.17 R_\odot$ while \cite{SK1970} model value is 25\% at a distance of $1.16 R_\odot$. Although the maximum distances of the model and observation are relatively close to each other, there is a significant difference of 11\% between the two polarisation degrees. On the other hand, the maximum values of this observation appear to be close to the maximum values of \cite{KH1958} in both regions.

\subsection{K coronal brightness and Its Polarization}\label{sec:41}
\begin{figure}[h!]
	\begin{minipage}[b]{0.48\textwidth}
		\centerline{\includegraphics[width=0.99\textwidth]{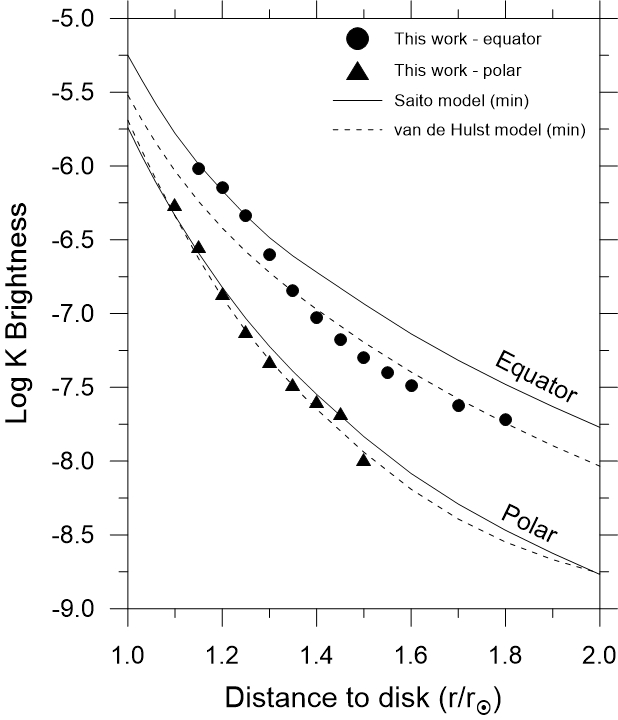}}
		\captionsetup[figure]{labelsep=space}
		\captionof{figure}{The $K$ corona values of the 2006 eclipse compared with the models of \protect\cite{VDH1950} and \protect\cite{SK1970}.} \label{fig:9}
	\end{minipage}
	\hspace{10pt}
	\begin{minipage}[b]{0.48\textwidth}  
		\centerline{\includegraphics[width=0.99\linewidth]{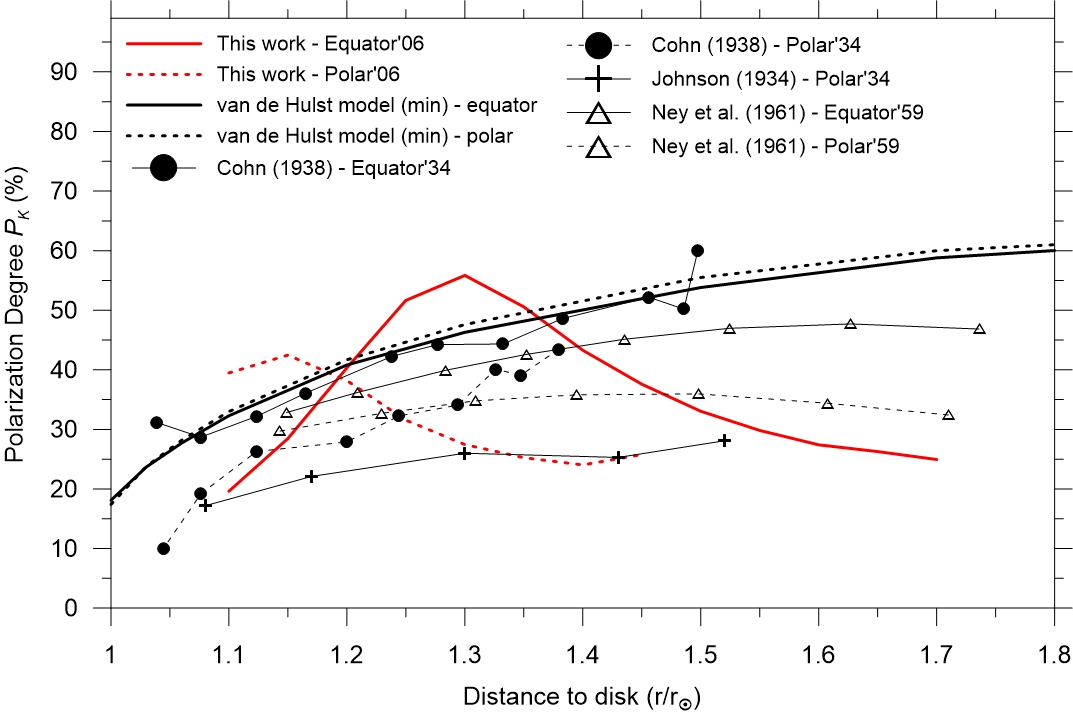}}
		\caption{The polarization degree $P_{\rm K}$ values of the $K$ corona (solid red lines) in comparison with the model values of \protect\cite{VDH1950} (black solid lines) and observational results obtained by other authors (lines with symbol).}
	     \label{fig:10}
	\end{minipage}
\end{figure}

Several methods are described in the literature for separating the $K$ and $F$ corona contributions to the total corona \citep{VDH1950, KH1958, NEP1961, KS1978}. The methods used by \cite{KH1958} and \cite{KS1978} have been applied separately in this study. The $F$ corona values obtained with these two methods were then compared with the $F$ model values of \cite{VDH1950}. However, as shown in Figure~\ref{fig:8}, the obtained $F$ corona values for the 2006 eclipse are not sufficiently satisfactory in comparison to the model values. In particular, the $F$ corona values before a distance of  1.4$R_\odot$ appear to be elevated and affected by the distribution of coronal matter formed by the present Solar Cycle. Therefore, a general approach has been adopted in light of the following explanations. The studies conducted by \cite{VDH1950}, \cite{SK1977}, \cite{MH2007}, and \cite{HY2012} provide evidence that the $F$ corona remains relatively stable from cycle to cycle. According to this approach, it is assumed that the $F$ coronal brightness, which consists of light reflected from planetary dust around the Sun, does not change much. As a result, the model values of the $F$ corona obtained from previous observations can be used in all other eclipse observations. Therefore, in this study, $K$ coronal brightness of the 2006 eclipse is calculated by subtracting the $F$ corona model values of \cite{VDH1950} from the observational total coronal $K+F$ brightness of this eclipse. The values of $K$ coronal brightness obtained are presented in Table~\ref{tab:4} for both the equatorial and polar regions. These values are also illustrated in Figure~\ref{fig:9}, which compares them with the model values of \cite{VDH1950} and \cite{SK1970}.

The degree of polarization $P_{\rm K}$ of the $K$ corona is calculated using the definition given in \cite{KS1978} as
\begin{equation}\label{equ:7}
	P_{\rm K} = \bigg( \frac{K + F}{K} \bigg) \, P_{\rm K + F}
\end{equation}
where ${K+F}$ and $P_{\rm K+F}$ are the total coronal brightness and its degree of polarization, respectively. The calculated polarization degree $P_{\rm K}$ values of the $K$ corona are shown in Figure~\ref{fig:10} in comparison with both equatorial and polar model values of \cite{VDH1950} and observational results obtained by several authors. The maximum value obtained for the equatorial region is 55\% at a distance of $1.3R_\odot$, and for the polar region, it is 42\% at a distance of $1.15R_\odot$.

As can be seen in Figure~\ref{fig:10}, these maximum values of polarization degree are slightly higher than the other values. This discrepancy may be attributed to the asymmetric distribution of coronal structures, which varies with the phase of the solar cycle. Such asymmetry can lead to an increased accumulation of coronal features at certain heliocentric distances. Additionally, a sharp decline in the degree of polarization is observed following the peak values. This rapid decrease is likely due to the absence of long-exposure images exceeding 1 second. In the present study, all exposures were shorter than 1 second, and the telescope used—featuring an 8-inch aperture—was relatively wide for this type of eclipse observation. Consequently, the solar coronal brightness may have been slightly overestimated at specific distances. This interpretation is also supported by the definition of the degree of polarization, which is given by
\begin{equation}\label{equ:8}
	P_{\rm K} = \frac{K_{\rm t} - K_{\rm r}}{K_{\rm t} + K_{\rm r}} = \frac{K_{\rm t} - K_{\rm r}}{K}
\end{equation}
where  $K_{\rm t}$ and $K_{\rm r}$ are the tangential and radial component of the $K$ corona light, respectively \citep{SA1879,BS1939,VDH1950}. According to this definition, the polarization degree is inversely proportional to the $K$ coronal brightness. Therefore, an overestimation of coronal brightness leads to a lower calculated polarization degree. Had longer exposure times been employed, the average $K$-coronal brightness values of the would likely have been slightly lower. As a result, the polarization degree values beyond certain radial distances would have been more closely matched by the expected values.
\section{Electron Density}\label{sec:5}
According to the corona models developed to date, the $K$ corona light is caused by the scattering of free electrons. Therefore, the $K$ coronal brightness is directly related to the number of electrons along the line of sight \citep{KH1958, KK2005}. With this perspective, \cite{VDH1950} has developed a method using successive approximations to calculate electron density using the observed $K$ coronal brightness. With this approach, the author used the following equations;
\begin{align}
	K_{\rm t} &= \sum_s h_{\rm s} \, x^{\rm -s} = 1/2 \, (1 + P_{\rm K}) \, K \label{equ:9}
\end{align}
\begin{align}
	K_{\rm t} - K_{\rm r} &= \sum_{\rm s} k_{\rm s} \, x^{\rm -s} = P_{\rm K} \, K\label{equ:10}
\end{align}
where $\sum_{\rm s} h_{\rm s} \, x^{\rm -s}$ is a power series which is written as a three-element polynomial as $A x^{\rm -a} + B x^{\rm -b} + C x^{\rm -c}$. And other equations;\vspace*{-4pt}
\begin{alignat}{2}\label{equ:11}
r\, C\, N(r)\, A (r)  &= \sum_{\rm s} \frac{h_{\rm s}}{a_{\rm s-1}} \, r^{\rm -s} && = K_{\rm t} (r) 
\end{alignat}\vspace*{-10pt}
\begin{alignat}{2}\label{equ:12}
	r\, C\, N(r) \big\{ A(r) - B(r) \big\} &= \sum_{\rm s} \frac{k_{\rm s}}{a_{\rm s+1}} \, r^{\rm -s} && = K_{\rm t} (r) - K_{\rm r} (r)
\end{alignat}
where $a_{\rm s}$ is a coefficient which calculated using gamma function given by author, $C$ is a coefficient which equals to $3.44 \times 10^{-14} $cm$^3$ and, $A$ and $B$ are the length of semi-major and semi-minor axis of the vibration ellipsoid, respectively (please refer to the article by \cite{VDH1950} to calculate $A$ and $B$ values depending on the distance to the solar disk). The method developed by van de Hulst is somewhat difficult to apply, and many more attempts need to be made.

At this point, to simplify the calculations and reduce the number of iterations, the \cite{VDH1950} equations given in Equations(\ref{equ:11} and \ref{equ:12}) were mathematically rearranged from a different perspective, and a new calculation method was developed to obtain the corona electron density (for more information, see the article by \cite{HC2023}). According to this method, it is possible to define two different electron densities by
\begin{alignat}{2}
	N_{\rm t} (r) &= \frac{f[K_{\rm t} (r)]}{r\, C\, A (r)} \label{equ:13}\\[1.5ex]
    N_{\rm t-r}(r) &= \frac{f[K_{\rm t} (r) - K_{\rm r} (r)]}{r\, C\, \big\{ A(r) - B(r) \big\}}\label{equ:14}
\end{alignat}
where $N_{\rm t} (r)$ and $N_{\rm t-r}(r)$ are the electron densities, and $f[K_{\rm t} (r)]$ and $f[K_{\rm t} (r) - K_{\rm r} (r)]$ are the generated functions (GFs) for the $K_{\rm t} (r)$ and  $K_{\rm t} (r) - K_{\rm r} (r)$ components, respectively. Since the coronal density $K$ is linearly proportional to the electron density $N(r)$, the relationship $K=2 \, K_{\rm t} - (K_{\rm t} - K_{\rm r})$ obtained from Equations(\ref{equ:9} \& \ref{equ:10}) and valid between the corona components $K_{\rm t} (r)$ and  $K_{\rm t} (r) - K_{\rm r} (r)$  should also be valid for the electron density components. Therefore, the total electron density, $N(r)$, can be expressed similarly in terms of its components as follows:
\begin{equation}\label{equ:15}
	N = 2 \, N_{ \rm t}  - N_{\rm t - r}
\end{equation}
Returning to the calculations, the $K$ coronal brightness and polarisation degree $P_{\rm K}$ have been appropriately calculated up to this stage. Now, Equations(\ref{equ:9} \& \ref{equ:10}) can be used to calculate the $K$ corona components $K_{\rm t} (r)$ and $K_{\rm t} (r) - K_{\rm r} (r)$ separately. By subsequently fitting $A x^{\rm -a} + B x^{\rm -b} + C x^{\rm -c}$ form polynomial curves to the values of these components, the coefficients $s$, $h_{\rm s}$ and $k_{\rm s}$, as specified in Equations(\ref{equ:9} and \ref{equ:10}), can be obtained (see the article by \cite{HC2023} for details). After the new coefficients of $h_{\rm s}/a_{\rm s-1}$ and $k_{\rm s}/a_{\rm s+1}$ been calculated for each component of $K_{\rm t}$ and $K_{\rm t} - K_{\rm r}$ shown in Equations(\ref{equ:11} and \ref{equ:12}), the GFs are formed using these new coefficients by
\begin{equation}\label{equ:16}
	\begin{split}
	    f[K_{\rm t} (r)] &= \sum_s \frac{h_{\rm s}}{a_{\rm s-1}} \, r^{\rm -s} \\[1ex] 
		f[K_{\rm t} (r) - K_{\rm r} (r)] &= \sum_s \frac{k_{\rm s}}{a_{\rm s+1}} \, r^{\rm -s} 
	\end{split}
\end{equation}
The electron densities $N_{\rm t} (r)$ and $N_{\rm t-r}(r)$ of $K_{\rm t}$ and $K_{\rm t} - K_{\rm r}$ components are then calculated by using Equations(\ref{equ:13} and \ref{equ:14}). Finally, the total electron density required for the $K$ coronal brightness is calculated by using Equation(\ref{equ:15}).

\begin{figure}[h!]    
	\centerline{\includegraphics[width=.45\linewidth]{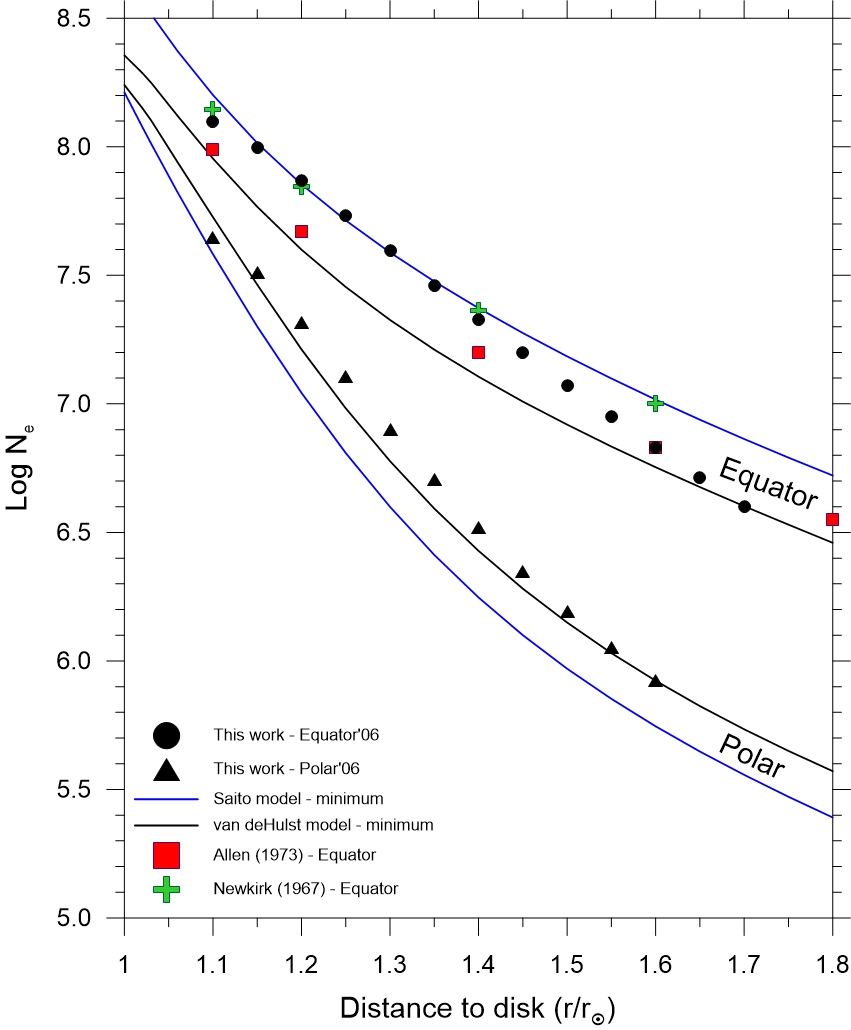}}
	\caption{The electron density values for the 2006 eclipse compared with the minimum corona model values \protect\cite{VDH1950} and \protect\cite{SK1970} and several observational results.}
    \label{fig:11}
\end{figure}

During the calculations, radial directions with polar angles between 260$^\circ$ and 280$^\circ$ and 315$^\circ$, where active regions or powerful coronal streamers exist are not taken into account and excluded from the computations. The calculated electron densities for the 2006 eclipse in both the equatorial and polar regions are listed in the rightmost column of Table~\ref{tab:5}, together with the other parameters used in the calculating process. These electron densities are also shown graphically in Figure~\ref{fig:11} compared with the minimum corona model values of \cite{VDH1950} and \cite{SK1970} and the observational results of \cite{ACW1973} and \cite{NDS1967}. As can be seen in Figure~\ref{fig:11}, the electron density values obtained for the 2006 eclipse are in good agreement with those of all others. However, when these values are compared with those of the models, the observed differences in electron density depending on the distance from the disk edge are interpreted as resulting from the asymmetric distribution of matter around the solar disk, which is specific to the 2006 eclipse.
\begin{table*}[h!]
	\setlength{\tabcolsep}{5.2pt}
	\renewcommand{\arraystretch}{0.6}
	\setlength{\extrarowheight}{7.5pt}\small\centering
	\caption{The values used to calculate electron density (left side) and the results obtained for the 2006 eclipse (right side) are shown for both the equatorial and polar regions.
	}
	\begin{tabular}{crrrr|rrrrr}
    \hline
    \hline
		\multicolumn{10}{c}{$A$-equatorial region} \\ 
		\cline{1-10} 
		$r$ & \multicolumn{1}{c}{$K$}& \multicolumn{1}{c}{$P_{\rm K}$}
		& $K_{\rm t}-K_{\rm r}$ & \multicolumn{1}{c|}{$K_{\rm t}$}
		& $f(K_{\rm t}-K_{\rm r})$ & $f(K_{\rm t})$ & $N_{\rm t-r}$ 
		& \multicolumn{1}{c}{$N_{\rm t}$} & \multicolumn{1}{c}{$N_{\rm e}$} \\
		\cline{1-10}  
		1.10 & 158.55 & 0.196 & 31.10 & 94.82 & 160.47 & 269.28 & 197.19 & 161.28 & 125.4 \\
		1.15 &  96.57 & 0.285 & 27.49 & 62.03 & 103.45 & 171.48 & 118.95 & 109.00 &  99.0 \\
		1.20 &  71.62 & 0.403 & 28.89 & 50.25 &  67.98 & 111.41 &  75.39 &  74.62 &  73.9 \\
		1.25 &  45.99 & 0.517 & 23.75 & 34.87 &  45.48 &  73.74 &  49.49 &  51.79 &  54.1 \\
		1.30 &  24.91 & 0.558 & 13.91 & 19.41 &  30.92 &  49.64 &  33.38 &  36.44 &  39.5 \\
		1.35 &  14.39 & 0.506 &  7.28 & 10.83 &  21.35 &  33.95 &  23.03 &  25.97 &  28.9 \\
		1.40 &   9.39 & 0.433 &  4.06 &  6.73 &  14.94 &  23.56 &  16.19 &  18.74 &  21.3 \\
		1.45 &   6.64 & 0.376 &  2.50 &  4.57 &  10.60 &  16.57 &  11.58 &  13.68 &  15.8 \\
		1.50 &   4.99 & 0.330 &  1.65 &  3.32 &   7.61 &  11.80 &   8.40 &  10.10 &  11.8 \\
		1.55 &   3.95 & 0.298 &  1.18 &  2.56 &   5.52 &   8.51 &   6.18 &   7.53 &   8.9 \\
		1.60 &   3.24 & 0.274 &  0.89 &  2.06 &   4.05 &   6.20 &   4.60 &   5.67 &   6.8 \\
		1.65 &   2.39 & 0.263 &  0.63 &  1.51 &   3.00 &   4.56 &   3.46 &   4.31 &   5.2 \\
		1.70 &   1.92 & 0.249 &  0.48 &  1.20 &   2.25 &   3.39 &   2.63 &   3.31 &   4.0 \\
		\hline \hline
		\multicolumn{10}{c}{$B$-polar region} \\
		\hline \hline
		1.10 & 54.78 & 0.395 & 21.63 & 38.20 & 76.47 & 115.36& 93.97 & 69.09 & 44.2 \\
		1.15 & 28.49 & 0.425 & 12.10 & 20.30 & 35.99 & 58.04 & 41.39 & 36.89 & 32.4 \\
		1.20 & 13.68 & 0.382 & 5.22  & 9.45  & 18.31 & 30.60 & 20.30 & 20.49 & 20.7 \\
		1.25 & 7.56  & 0.316 & 2.39  & 4.97  & 10.05 & 16.90 & 10.93 & 11.87 & 12.8 \\
		1.30 & 4.71  & 0.275 & 1.29  & 3.00  & 5.92  & 9.78  & 6.39  & 7.18  &  8.0 \\
		1.35 & 3.29  & 0.252 & 0.83  & 2.06  & 3.71  & 5.92  & 4.00  & 4.53  &  5.1 \\
		1.40 & 2.52  & 0.240 & 0.61  & 1.56  & 2.45  & 3.74  & 2.65  & 2.98  &  3.3 \\
		1.45 & 2.10  & 0.258 & 0.54  & 1.32  & 1.68  & 2.46  & 1.84  & 2.03  &  2.2 \\
		1.50 & 1.03  & 0.287 & 0.29  & 0.66  & 1.19  & 1.68  & 1.32  & 1.44  &  1.6 \\
		1.55 & 0.50  & 0.266 & 0.13  & 0.32  & 0.87  & 1.18  & 0.97  & 1.05  &  1.1 \\
		1.60 & 0.34  & 0.264 & 0.09  & 0.22  & 0.64  & 0.86  & 0.73  & 0.79  &  0.8 \\
        \hline \hline
	\end{tabular}
	\label{tab:5}
    \end{table*}
\section{Average coronal Temperature in the Mid-Height}\label{sec:6}
Informative data on the temperature in the corona can be obtained using the \cite{VDH1950} approach, which is based on the assumption of hydrostatic equilibrium in the coronal atmosphere. According to this approach, hydrostatic equilibrium is formed by the sum of the gravitational force per unit volume and the pressure gradient. This condition is expressed as
\begin{equation}\label{equ:17}
	 \mu \, m_{\rm H} \, \frac{G\,M_\odot \,N_{\rm e}}{(r\,R_\odot)^2} + k \frac{d(N_{\rm e} \, T)}{R_\odot \, dr} = 0
\end{equation}
where $G$ is the universal gravitational constant, $R_\odot$ is the solar radius, $M_\odot$ is the mass of the Sun, $N_{\rm e}$ is the number of the electrons per cm$^3$, $\mu$ is the mean molecular weight of the coronal gas, and is taken as 0.61 \citep{NEP1961, KK2005} for this study, $m_{\rm H}$ is the mass of hydrogen atomic weight, $r$ is the distance from the Sun's center in terms of the solar radius, $k$ is the Stefan-Boltzmann constant and, $T$ is the temperature \citep{VDH1950}. When this equation is arranged, the following form will be obtained:
\begin{equation}\label{equ:18}
	\frac{T_1 \, N_{\rm e}}{r^2} = - \frac{d(N_{\rm e} \, T)}{dr}
\end{equation}
where $T_1 = \mu \, m_{\rm H} \, G\, M_\odot /R_\odot \, k  = 14.1 \times 10^6$ K. Finally, this equation can easily be converted to the form 
\begin{equation}\label{equ:19}
	\frac{T_{\rm 1}}{T} = \frac{d}{d(1/r)} \; \text{ln} \, N_{\rm e} + \frac{d}{d(1/r)} \; \text{ln} \, T
\end{equation}
In this equation, the first term denotes the gradient of electron density relative to distance, while the second term indicates the gradient of temperature. It is generally accepted that, under the assumption of hydrostatic equilibrium, this second term is negligible in the solar corona \citep{VDH1950}. Due to he gradient of temperature being negligible, with minor fluctuations occurring over comparatively brief distances within the corona, the determination of the temperature can be achieved from the gradient of the linear regression line (ln $N_{\rm e}$ versus $1/r$) derived from the initial term of Equation(\ref{equ:20}) as
\begin{equation}\label{equ:20}
	\frac{T_{\rm 1}}{T} = \frac{d}{d(1/r)} \; \text{ln} \, N_{\rm e}
\end{equation}

\begin{figure}[t!]    
	\centerline{\includegraphics[width=0.65\linewidth]{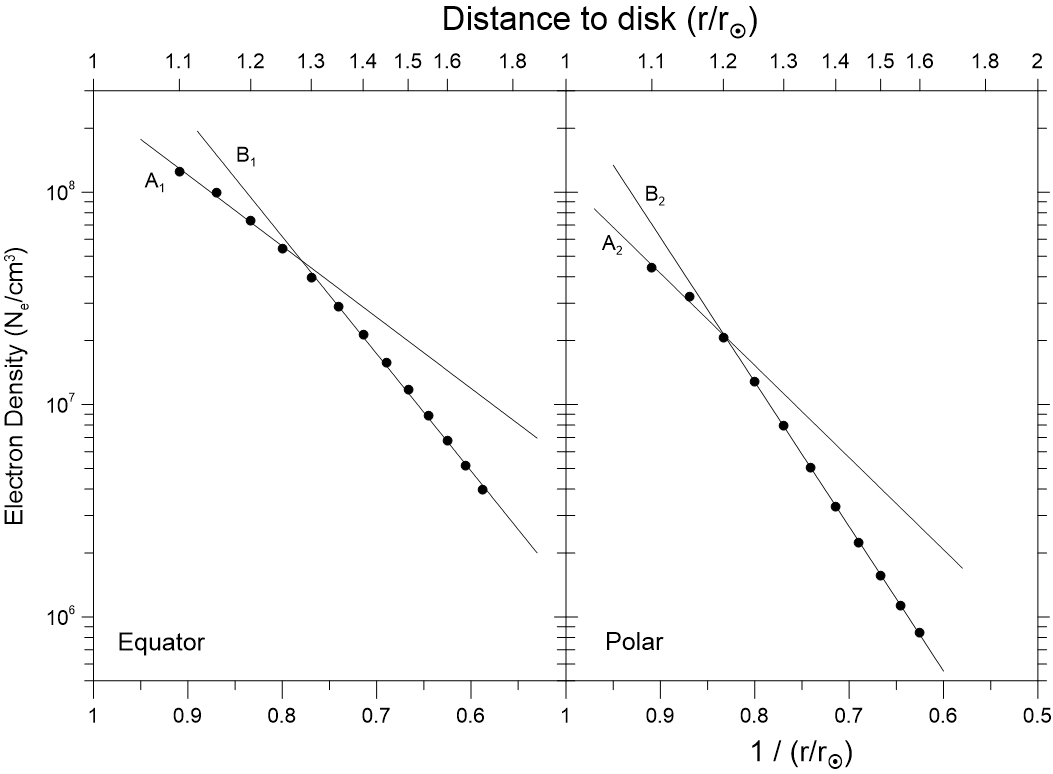}}
	\caption{The graph of electron density gradient as a function of disk distance in the equatorial and polar regions. Solid lines indicate the range where a linear gradient occurs.}
    \label{fig:12}
\end{figure}

The electron density versus $1/r$ plots for the 2006 eclipse in both the equatorial and polar regions are presented in Figure~\ref{fig:12}. As illustrated in the figure, the linear relationship predicted for an isothermal corona is not generally valid, and two distinct regions, designated as the near and far areas to the disk, are observed. After identifying areas showing a relatively linear slope in the electron density graph, the lines are adjusted to fit these areas. The temperature of these regions is then obtained by employing the slopes of these straight lines, as outlined in Equation~\ref{equ:20}. The equations of the fitted lines for both the equatorial and polar regions of the 2006 eclipse are separately listed in Table~\ref{tab:6} for the near ($A_{\rm 1}$, $A_{\rm 2}$) and far ($B_{\rm 1}$, $B_{\rm 2}$) areas of the disk, together with the calculated temperature values.

The temperature values calculated for the equatorial region during the 2006 eclipse were 1.82$\times 10^6$ K at a distance of 1.2$R_\odot$ and 1.11$\times 10^6$ K at a distance of 1.5$R_\odot$, and for the polar region, 1.15$R_\odot$ and 0.82$\times 10^6$ K at a distance of 1.4$R_\odot$, are slightly higher than the values obtained by both \cite{NEP1961} and \cite{KK2005}. In the eclipse of 2 October 1959, \cite{NEP1961} obtained the temperature values as 1.22$\times 10^6$ K at a distance of 1.3$R_\odot$ in the equatorial region, and 0.91$\times 10^6$ K at a distance of 1.8$R_\odot$ in the polar region. In addition, during the eclipse of 11 August 1999, \cite{KK2005} obtained temperature values of 1.23$\times 10^6$ K at a distance of 1.3$R_\odot$ in the equatorial region and 1.07$\times 10^6$ K at a distance of 1.8$R_\odot$ in the polar region.

One potential explanation for the observed variations in the 2006 eclipse may be attributed to the 23rd Solar Cycle, which exhibited characteristics that differed from those of the preceding cycle. This cycle was characterised by an increased prevalence of sunspots with larger areas, the formation of brighter faculae and plage areas \citep{TG2004, KA2011}, and a higher frequency of coronal mass ejections \citep{KA2011a}. In contrast, the obtained temperature values demonstrate a high degree of consistency with the values of \cite{VDH1950}, which are reported to be 1.62$\times 10^6$ K in the equatorial region and 1.15$\times 10^6$ K in the polar region, for distances ranging from 1 to 3$R_\odot$.
\begin{table}[h!]
	\setlength{\tabcolsep}{5pt}
	\renewcommand{\arraystretch}{1.8}
	\small\centering
	\caption{The linear equations of the electron density gradient in the equator and polar region for the 2006 eclipse, and their mean distance and temperature values.}\vspace*{3pt}
	\begin{tabular}{cclcc}
        \cline{1-5}
		&  & \multicolumn{1}{l}{\multirow{2}[-2]{*}{Equations of the straight lines}} 
		& $\langle\,r\,\rangle$ & $\langle\,T\,\rangle$ \\[-2ex]
		& & & ($R_\odot$) & ($\times$10$^6$K) \\
		\cline{2-5}
		\multicolumn{1}{c}{\multirow{2}[3]{*}{\rot{\parbox[t]{1.3cm}{\hspace*{8pt}Equator}}}}  
		& $A_1$ & $\ln N_{\rm e} = ~7.71381\times (1/r) + 11.66555$   & 1.2   & 1.82 \\[-1.5ex]
		& $B_1$ & $\ln N_{\rm e} = 12.70187\times (1/r) + ~7.77761$  & 1.5   & 1.11 \\[0.6ex]
		\cline{2-5}
		\multicolumn{1}{c}{\multirow{2}[3]{*}{\rot{\parbox[t]{1.5cm}{\hspace*{16pt}Polar}}}}
		& $A_2$ & $\ln N_{\rm e} = ~~9.98710\times (1/r) + 8.55229$ & 1.2   & 1.41 \\[-1.5ex]
		& $B_2$ & $\ln N_{\rm e} = 15.66973\times (1/r) + 3.82678$ & 1.4   & 0.89 \\[0.6ex]
		\cline{1-5}
	\end{tabular}
\label{tab:6}
\end{table}

\section{Conclusion}\label{sec:7}
The physical parameters obtained from the 2006 eclipse are generally consistent with both the minimum corona model values of \cite{VDH1950} and \cite{SK1970} and some observational results in the literature. In particular, the total coronal ($K+F$), and $K$ brightness, electron density, and average coronal temperature values are in good agreement with the model values and those of others. However, no such similar agreement is observed for the polarisation degree values. As previously explained, the possible reasons for this phenomenon include the use of a relatively wide-aperture telescope and, more importantly, the absence of exposures longer than 1 second, as well as the asymmetric distribution of coronal brightness around the solar disk.  

The fact that the results of the eclipse on 29 March 2006 are generally consistent with model values is primarily because the 2006 eclipse occurred during the minimum period of the 23rd Solar Cycle, and the new method used in this observation is based on a correct approach. This observational consistency is also of particular importance in confirming the accuracy of the developed models in some respects. For instance, the variation in brightness with distance is generally consistent with the model's predictions. However, it should be noted that these are not identical, and minor discrepancies may be observed, which vary from cycle to cycle depending on the activity of the Solar Cycle. The observed differences indicate that the matter in the corona does not have an isotropic distribution, as previously accepted by \citep{NEP1961}. Consequently, it would be beneficial to consider modifying some of the approaches in current models for evaluation from a different perspective.

The method developed for the density calibration of roll films represents a novel approach in this field. Therefore, one of the most significant phases of this study must be conducted with the greatest care. In this instance, images of the solar disk captured with filters and varying aperture sizes are employed to derive the density calibration function of the film utilised in the observational process. A careful examination of Figure~\ref{fig:4} reveals that short exposure values are located in the lower left part of the curve, while long exposure values are located in the upper right part. The curvature is dependent on the exposure times used. Consequently, the range of exposure times to be utilised for the accurate creation of this calibration curve must be carefully selected. A significant feature of this function is to provide the contributions of the sky and instrumental intensity ($I_{\rm A}+I_{\rm S}$) to the measured total intensity. It is important to note that, in the context of film intensity measurement, the intensity value is typically normalised against the background intensity, which is represented by the sky. Consequently, the normalised intensity value of 1 is equivalent to the intensity of the sky, inclusive of any additional contributions. Furthermore, it is important to note that the parameters used to create this function are observational parameters, including disk images, filters, exposure, and apertures. In this regard, it can be stated that the density calibration function obtained from these parameters is also an observational result.

The second method developed in this study is the direct calculation of electron density without using any approach method similar to those applied by \cite{VDH1950} and \cite{KH1958}. It can be said that the developed method is a relatively practical and easy-to-use approach compared to alternative methods. As demonstrated in Figure~\ref{fig:11}, the observation of values that correspond to those of the model suggests that the developed method exhibits a comparatively precise approach. However, it is essential to consider a critical point regarding the method. In the process of polynomial curve fitting, it is important to ensure that the fitted curve passes through the general distribution of the observation points rather than passing close to each observation point. This is due to the presence of observational errors, which lead to deviations in the observed values.

While the results obtained for the 2006 eclipse are satisfactory, it would be beneficial to retest these new methods with other eclipse observations. Consequently, it would be valuable to establish contact with multiple researchers working on this subject in the future.

\begin{description}
    \item[Peer Review:] Externally peer-reviewed.
    \item[Author Contribution:] Conception/Design of study - H.\c{C}.;~ Data Analysis/Interpretation - H.\c{C}.;~Drafting Manuscript - H.\c{C}.;~Critical Revision of Manuscript - H.\c{C}.;~Final Approval and Accountability - H.\c{C}.
    \item[Conflict of Interest:] Author declared no conflict of interest.
    \item[Financial Disclosure:] This work was supported by the Istanbul University Scientific Research Projects Commission with project numbers 24242 and 470/27122005.
\end{description}

\section*{Acknowledgements}
 Thanks to every staff member of the Department of Astronomy and Space Sciences who participated in the 2006 solar eclipse observation. Thanks to the anonymous referee for their valuable suggestions and comments, which significantly enhanced the manuscript.
\bibliographystyle{mnras}
\bibliography{refsolecl}

\bsp	
\label{lastpage}
\end{document}